\documentclass[12pt,a4paper]{article}
\pdfoutput=1
\usepackage[a4paper]{geometry}
\usepackage{graphicx}
\usepackage{amsmath}
\usepackage{amssymb}
\usepackage{float}
\usepackage{listings}
\newcommand{\asb}{\bar{\alpha}_s}

\newcommand{\stringa}{\ttfamily\lstinline}
\def\cod#1{{\stringa!#1!}}

\title{\bf Inclusive Four-jet Production  at 7 and 13 TeV: Azimuthal Profile in Multi-Regge Kinematics}
\author{F. Caporale$^1$, F.~G. Celiberto$^{1,2}$, G. Chachamis$^1$, \\ 
        D. Gordo G{\' o}mez$^1$\footnote{`la Caixa'-Severo Ochoa Scholar.}\,\,, A. Sabio Vera$^{1}$\\ \\
{\small $^1$ Instituto de F{\' \i}sica Te{\' o}rica UAM/CSIC, Nicol{\'a}s Cabrera 15}\\ 
{\small \& Universidad Aut{\' o}noma de Madrid, E-28049 Madrid, Spain.}\\
{\small $^2$ Dipartimento di Fisica, Universit{\`a} della Calabria \&}\\
{\small Istituto Nazionale di Fisica Nucleare, Gruppo Collegato di Cosenza,}\\
{\small I-87036 Arcavacata di Rende, Cosenza, Italy.}
}

\begin{document}

\maketitle 

\abstract
Recently, new observables in LHC inclusive events with three tagged jets were proposed. Here, we extend that proposal to events with four tagged jets. The events are characterised by one jet in the forward direction, one in the backward direction with a large rapidity distance $Y$  from the first one and two more jets tagged in more central regions of the detector. In our setup, non-tagged associated mini-jet multiplicity is present and needs to be accounted for by the inclusion of BFKL gluon Green functions. The projection of the cross section on azimuthal-angle components opens up the opportunity for defining new ratios of correlation functions of the azimuthal angle differences among the tagged jets that can be used as probes of the BFKL dynamics.

\section{Introduction}

The Large Hadron Collider (LHC) gives a unique opportunity to study high energy scattering in Quantum Chromodynamics (QCD). Jet production studies play a crucial role in high energy QCD phenomenology since the plethora of data makes possible the analysis of even more exclusive observables than usual. Here, we concentrate on four jet production in the so-called multi-Regge kinematics. In an experimental setup containing final state jets with a large rapidity separation, the Balitsky-Fadin-Kuraev-Lipatov (BFKL) framework in the leading logarithmic (LL)~\cite{Lipatov:1985uk,Balitsky:1978ic,Kuraev:1977fs,Kuraev:1976ge,Lipatov:1976zz,Fadin:1975cb} and next-to-leading logarithmic (NLL) approximation~\cite{Fadin:1998py,Ciafaloni:1998gs} presents itself as a powerful tool to probe the dominant dynamics of the QCD high energy limit. 
\begin{figure}[H]
 \centering
 \includegraphics[scale=0.5]{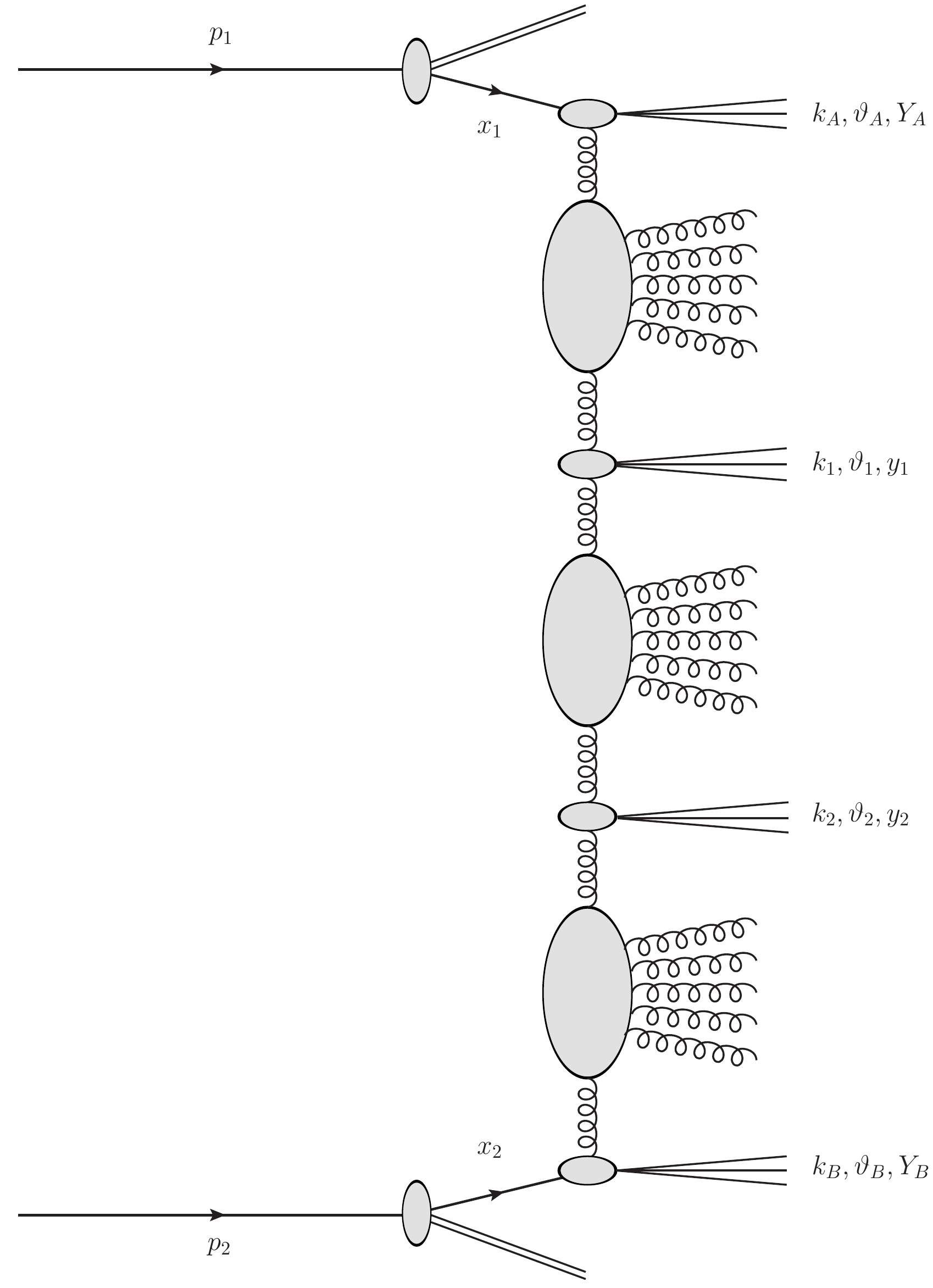}
 \caption[]
 {Inclusive four-jet production process in multi-Regge kinematics.}
 \label{fig:4j}
 \end{figure}Arguably, jet production studies in the last decade searching for the onset of BFKL effects
 were mainly focused  on
 Mueller-Navelet jets (dijets)~\cite{Mueller:1986ey}  in hadronic colliders. They correspond to the inclusive production of two jets with similar transverse sizes, $k_{A,B}$,  and a large rapidity difference $Y=\ln ( x_1 x_2 s/(k_A k_B))$. $x_{1,2}$ are the usual Bjorken parameters of the two partons 
that are linked to the jets and $s$ is the centre-of-mass-energy squared. 
A number of analyses~\cite{DelDuca:1993mn,Stirling:1994he,Orr:1997im,Kwiecinski:2001nh,Angioni:2011wj,Caporale:2013uva, Caporale:2013sc}  of the average values, $\langle \cos{(m \, \phi)} \rangle$, for the azimuthal-angle difference between the two tagged jets, $\phi$, suggest minijet activity in the rapidity interval between the most forward and most backward jet
which cannot be dismissed and which affects the azimuthal angle difference of these jets.
A downside is that collinear effects~\cite{Vera:2006un,Vera:2007kn}, having their origin at the zeroth component of the conformal spin, affect significantly the azimuthal 
angle observables. Nevertheless,
this collinear contamination can be mostly eliminated if the ratios of projections on azimuthal-angle observables~${\cal R}^m_n = \langle \cos{(m \, \phi)} \rangle / \langle \cos{(n \, \phi)} \rangle$~\cite{Vera:2006un,Vera:2007kn} (where $m,n$ are integers and $\phi$ the azimuthal angle between the two tagged jets) are considered instead. Moreover, the ratios offer a more clear  signal of BFKL effects than the standard predictions for the growth of hadron structure functions $F_{2,L}$ (well fitted within  NLL approaches~\cite{Hentschinski:2012kr,Hentschinski:2013id}). The confrontation of different NLL theoretical predictions for these ratios ${\cal R}^m_n$~\cite{Marquet:2007xx,Ducloue:2013bva,Caporale:2014gpa,Caporale:2014blm,Celiberto:2015dgl,Celiberto:2016ygs,Colferai:2010wu,Ducloue:2013wmi,Ducloue:2014koa,Aad:2014pua,Khachatryan:2016udy,Mueller:2015ael,Chachamis:2015crx,N.Cartiglia:2015gve} 
against LHC experimental data has been quite successful. 

In Refs.~\cite{Caporale:2015vya,Caporale:2016soq}, we
proposed new observables associated to the inclusive production of three jets. 
We argued there that the new observables feature 
appealing attributes.
Firstly, in order to obtain data for the proposed
three-jet analysis, one would only need to
search for an additional central jet within any existing 
Mueller-Navelet analysis data set. Secondly, 
on more theoretical grounds,
these observables probe fundamental characteristics of the BFKL ladder.
They correspond
to the ratios 
\begin{eqnarray}
{R}^{M N}_{P Q} =\frac{ \langle \cos{(M \, \phi_1)} \cos{(N \, \phi_2)}  \rangle}{\langle \cos{(P \, \phi_1)} \cos{(Q \, \phi_2)} \rangle} \, , 
\label{Rmnpq}
\end{eqnarray}
where $\phi_1 = \vartheta_A - \vartheta_J - \pi$ and $\phi_2 =  \vartheta_J - \vartheta_B - \pi$ 
with $\vartheta_{A,J,B}$ being the azimuthal angle of the forward, central and backward jet
respectively.

Here, we extend  our discussion to the case of four-jet events (different experimental analyses 
can be found in Refs.~\cite{Chatrchyan:2013qza,Aad:2015nda,Aaboud:2016dea}). For the present study, 
we need to have
one jet in the forward direction with rapidity $Y_A$, one in the backward direction with rapidity $Y_B$ and both well-separated in rapidity from the each other
so that $Y_A -Y_B$ is large, along with two extra jets tagged in more central regions of the detector. Additionally, the relative rapidity separation between any two neighbouring jets cannot be very different than
one third of $Y_A -Y_B$ so that the kinematical configurations
of the events actually follow the multi-Regge kinematics.
Extending Eq.~(\ref{Rmnpq}) to the partonic four-jet production, 
we studied in~\cite{Caporale:2015int} different ratios of three cosines in numerator and denominator:
\begin{eqnarray}
{\cal R}^{M N L}_{P Q R} =\frac{ \langle \cos{(M \, \phi_1)} \cos{(N \, \phi_2)} \cos{(L \, \phi_3)} \rangle}{\langle \cos{(P \, \phi_1)} \cos{(Q \, \phi_2)} \cos{(R \, \phi_3)} \rangle} \, , 
\label{Rmnlpqr}
\end{eqnarray}
where $\phi_1$, $\phi_2$ and $\phi_3$ are
 the azimuthal angle differences between neighbouring in rapidity jets. 
These ratios allow for the study of even more differential 
distributions in the transverse momenta, azimuthal angles and 
rapidities of the two central jets as well as for detailed work in connection to
multiple parton scattering~\cite{Jung:2011yt,Baranov:2015nma,Maciula:2015vza,Maciula:2014pla,Kutak:2016mik,Kutak:2016ukc,Ducloue:2015jba}.

In this paper, we define and study ratios of three cosines in numerator and denominator
beyond the partonic level. We make use of the collinear factorization scheme
to produce the two  uttermost jets and we convolute the partonic differential cross section, 
which is described by the BFKL dynamics, with collinear parton distribution functions.
We also include  in our computation the forward ``jet vertex"~\cite{Caporale:2012:IF,Fadin:2000:gIF,Fadin:2000:qIF,Bartels:2001ge,Bartels:2002yj}.
Three BFKL gluon Green functions link the outermost (Mueller-Navelet-like) jets with the 
more centrally produced ones. We integrate over the momenta of the four produced jets, 
using LHC kinematical cuts so that a
comparison of our predictions with forthcoming experimental analyses 
of LHC data is possible.
In the following section we will overview the main formulas
and  present our numerical results. We conclude with our Summary and Outlook.

 \begin{figure}[H]
 \centering
 \includegraphics[scale=0.55]{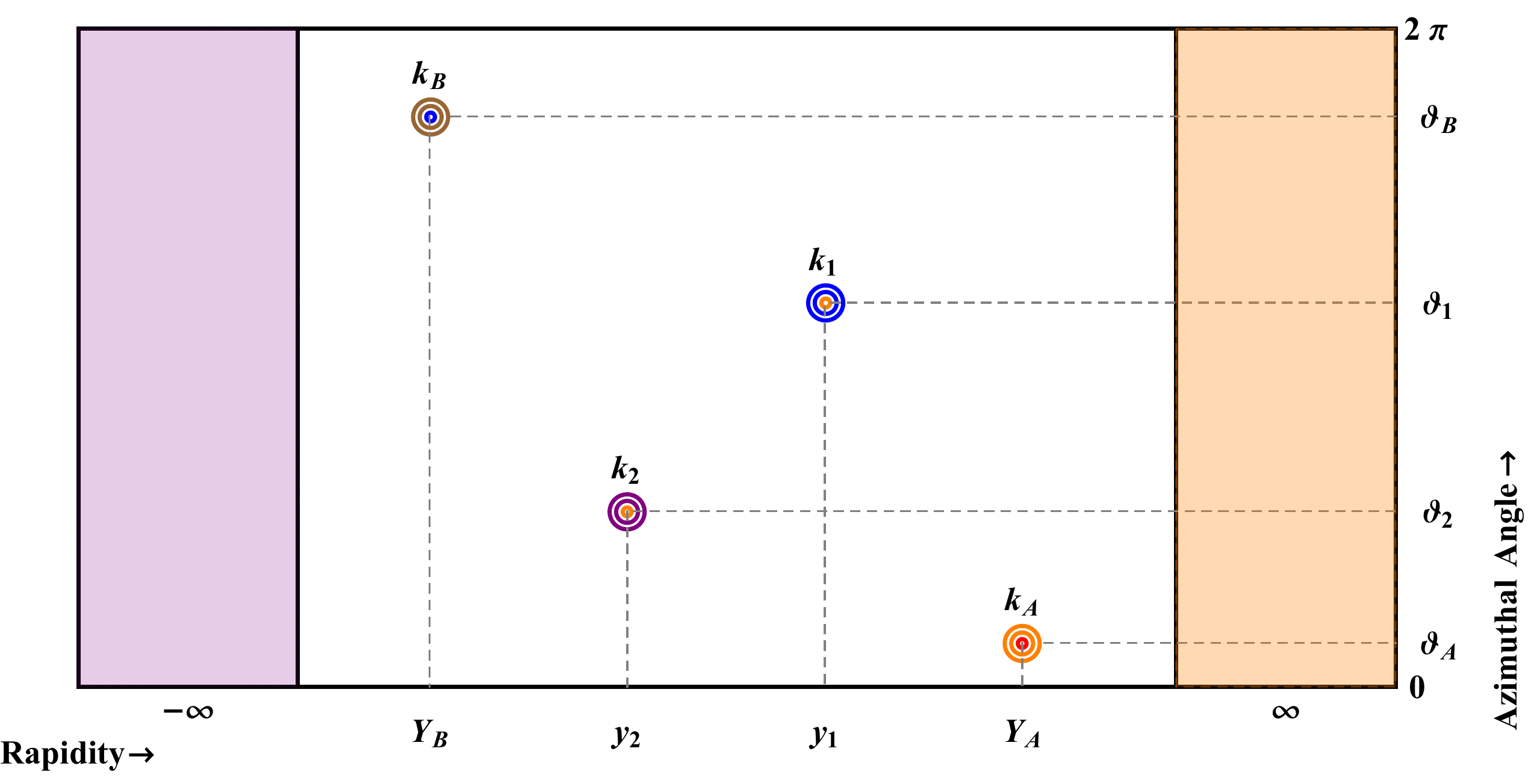}
 \caption[]
 {A primitive lego plot 
 depicting a four-jet event. $k_A$ is a forward jet with 
 large positive
 rapidity $Y_A$ and azimuthal angle $\vartheta_A$, $k_B$ is a backward jet with 
 large negative
 rapidity $Y_B$ and azimuthal angle $\vartheta_B$ and
 $k_1$ and $k_2$ are two jets with azimuthal angles $\vartheta_1$ 
 and $\vartheta_2$ respectively
 and rapidities $y_1$ and $y_2$
 such that $Y_A - y_1 \sim y_1- y_2 \sim y_2 - Y_B$. 
 }
 \label{fig:lego}
 \end{figure}

\section{Hadronic level inclusive four-jet production in multi-Regge kinematics}

We study (see Figs.~\ref{fig:4j} and~\ref{fig:lego})
the production of one forward and one
backward jet, both characterized by high transverse momenta $\vec{k}_{A,B}$ 
and well separated in rapidity, together with two more jets
produced in the 
central rapidity region and with possible associated mini-jet production: 
\begin{eqnarray}
\label{process}
{\rm proton }(p_1) + {\rm proton} (p_2) \to 
{\rm jet}(k_A) + {\rm jet}(k_1) + {\rm jet}(k_2) + {\rm jet}(k_B)  + {\rm minijets}
\end{eqnarray}

The cross section for the inclusive four-jet production process~(\ref{process}) reads
in collinear factorization
\begin{align}
\label{dsigma_pdf_convolution}
 & \frac{d\sigma^{4-{\rm jet}}}
      {dk_A \, dY_A \, d\vartheta_A \, 
       dk_B \, dY_B \, d\vartheta_B \, 
       dk_1 \, dy_1 d\vartheta_1  \, 
       dk_2 \, dy_2 d\vartheta_2}  = 
 \\ \nonumber 
 \hspace{1cm}& \sum_{\alpha,\beta=q,{\bar q},g}\int_0^1 dx_1 \int_0^1 dx_2
 \ f_{\alpha}\left(x_1,\mu_F\right)
 \ f_{\beta}\left(x_2,\mu_F\right) \;
 d{\hat\sigma}_{\alpha, \beta}\left(\hat{s},\mu_F\right) \;,
\end{align}
where $\alpha, \beta$ characterise the partons 
(gluon $g$; quarks $q = u, d, s, c, b$;
antiquarks $\bar q = \bar u, \bar d, \bar s, \bar c, \bar b$),
$f_{\alpha, \beta}\left(x, \mu_F \right)$ are the parton distribution functions of the protons; 
$x_{1,2}$ represent the longitudinal fractions of the partons involved 
in the hard subprocess; $d\hat\sigma_{\alpha, \beta}\left(\hat{s}, \mu_F \right)$ 
is the partonic cross section for the production of jets and
$\hat{s} \equiv x_1x_2s$ is the partonic squared center-of-mass energy 
(see Fig.~\ref{fig:4j}). The  cross-section 
for the partonic hard subprocess $d{\hat\sigma}_{\alpha, \beta}$
features a dependence on BFKL dynamics keeping in mind that
the emissions of mini-jet in the rapidity span between any two subsequent-in-rapidity jets
can be described by a forward gluon Green function $\varphi$. 

Making use of  the leading order approximation of the jet 
vertex~\cite{Caporale:2012:IF}, 
the cross section for the process~(\ref{process}) reads
\begin{align}
 & \frac{d\sigma^{4-{\rm jet}}}
      {dk_A \, dY_A \, d\vartheta_A \, 
       dk_B \, dY_B \, d\vartheta_B \, 
       dk_1 \, dy_1 d\vartheta_1 \, 
       dk_2 \, dy_2 d\vartheta_2} = 
 \nonumber \\ \hspace{1cm}&  
 \frac{16 \pi^4 \, C_F \, \asb^4}{N_C^3} \, 
 \frac{x_{J_A} \, x_{J_B}}{k_A \, k_B \, k_1\, k_2} \,
 \int d^2 \vec{p}_A \int d^2 \vec{p}_B  \int d^2 \vec{p}_1 \int d^2 \vec{p}_2 \,
 \nonumber \\  \hspace{1cm}& \times
 \delta^{(2)} \left(\vec{p}_A + \vec{k}_1- \vec{p}_1\right) 
 \delta^{(2)} \left(\vec{p}_B - \vec{k}_2- \vec{p}_2\right) \,
 \nonumber \\  \hspace{1cm}& \times 
 \left(\frac{N_C}{C_F}f_g(x_{J_A},\mu_F)
 +\sum_{r=q,\bar q}f_r(x_{J_A},\mu_F)\right) \,
 \nonumber \\ \hspace{1cm}& \times
 \left(\frac{N_C}{C_F}f_g(x_{J_B},\mu_F)
 +\sum_{s=q,\bar q}f_s(x_{J_B},\mu_F)\right)
 \nonumber \\ \hspace{1cm}& \times
 \varphi \left(\vec{k}_A,\vec{p}_A,Y_A - y_1\right) 
 \varphi \left(\vec{p}_1,\vec{p}_2,y_1 - y_2\right)
 %\nonumber \\ \hspace{1cm}& \times
 \varphi \left(\vec{p}_B,\vec{k}_B, y_2 - Y_B\right) .
\end{align}
In order to follow a multi-Regge kinematics setup, 
we demand that the rapidities of the produced particles 
obey $Y_A > y_1 > y_2 > Y_B$, while $k_1^2$ 
and $k_2^2$ are well
above the resolution scale of the detectors.
$x_{J_{A,B}}$ are the longitudinal momentum fractions
of the two external jets, connected to the respective rapidities 
$Y_{A,B}$ by the relation 
$x_{J_{A,B}} = k_{A,B} \, e^{\, \pm \, Y_{A,B}} / \sqrt{s}$.
The strong coupling is 
$\bar{\alpha}_s =   \alpha_s \left(\mu_R\right) \, N_c/\pi$ and
$\varphi$ are BFKL gluon Green functions following the normalization 
$ \varphi \left(\vec{p},\vec{q},0\right) = \delta^{(2)} \left(\vec{p} - \vec{q}\right)$.

Based upon the work presented 
in Refs.~\cite{Caporale:2016soq,Caporale:2015int}, 
we are after observables for which
the BFKL dynamics would surface in a distinct form. Moreover, we are interested
in observables that should be
rather insensitive to possible higher order corrections. Let us first define  the
following azimuthal angle differences:
$\phi_1 = \vartheta_A-\vartheta_1-\pi$,
$\phi_2 = \vartheta_1-\vartheta_2-\pi$,
$\phi_3 = \vartheta_2-\vartheta_B-\pi$.
Then, the related experimental observable we propose corresponds to
the mean value (with $M,N,L$ being positive integers)
\begin{align}
\label{Cmnl}
&{\cal C}_{MNL} \, = \,
 \langle  \cos(M \,\phi_1) \cos(N \,\phi_2)  \cos (L \,\phi_3 )
 \rangle = \nonumber \\
 & \frac{\int_0^{2 \pi} d \vartheta_A \int_0^{2 \pi} d \vartheta_B \int_0^{2 \pi} d \vartheta_1 
 \int_0^{2 \pi} d \vartheta_2 
 \cos(M \phi_1) \cos(N \phi_2)  \cos (L \phi_3 )
\; d \sigma^{4-{\rm jet}} }{\int_0^{2 \pi} d \vartheta_A d \vartheta_B d \vartheta_1 d \vartheta_2
 \; d \sigma^{4-{\rm jet}} }.
 \end{align}
The numerator in Eq.~(\ref{Cmnl}) actually reads
\begin{align}
\label{CmnlNUM}
& \int_0^{2 \pi} d \vartheta_A \int_0^{2 \pi} d \vartheta_B \int_0^{2 \pi} d \vartheta_1 
 \int_0^{2 \pi} d \vartheta_2  \cos(M \phi_1) \cos(N \phi_2)  \cos (L \phi_3 ) 
  \nonumber \\  \hspace{1cm}&
  \times \frac{d\sigma^{4-{\rm jet}}}
      {dk_A \, dY_A \, d\vartheta_A \, 
       dk_B \, dY_B \, d\vartheta_B \, 
       dk_1 \, dy_1 d\vartheta_1 \, 
       dk_2 \, dy_2 d\vartheta_2} =  
       \nonumber \\ &
 \frac{16 \pi^4 \, C_F \, \asb^4}{N_C^3} \, \frac{x_{J_A} \, x_{J_B}}{k_A \, k_B \, k_1\, k_2} \,
 \int d^2 \vec{p}_A \int d^2 \vec{p}_B  \int d^2 \vec{p}_1 \int d^2 \vec{p}_2 \,
 \nonumber \\  \hspace{1cm}& \times
 \delta^{(2)} \left(\vec{p}_A + \vec{k}_1- \vec{p}_1\right) 
 \delta^{(2)} \left(\vec{p}_B - \vec{k}_2- \vec{p}_2\right) \,
 \nonumber \\  \hspace{1cm}& \times 
 \left(\frac{N_C}{C_F}f_g(x_{J_A},\mu_F)
 +\sum_{r=q,\bar q}f_r(x_{J_A},\mu_F)\right) \,
 \nonumber \\ \hspace{1cm}& \times
 \left(\frac{N_C}{C_F}f_g(x_{J_B},\mu_F)
 +\sum_{s=q,\bar q}f_s(x_{J_B},\mu_F)\right)
 \nonumber \\ \hspace{1cm}& \times
 \left( 
 \tilde{\Omega}_{M,N,L} 
 + \tilde{\Omega}_{M,N,-L} 
 + \tilde{\Omega}_{M,-N,L} 
 + \tilde{\Omega}_{M,-N,-L} 
 \right.  
  \nonumber \\ \hspace{1cm}&
  \left.
\hspace{.5cm} + \tilde{\Omega}_{-M,N,L} 
 + \tilde{\Omega}_{-M,N,-L} 
 + \tilde{\Omega}_{-M,-N,L} 
 + \tilde{\Omega}_{-M,-N,-L} 
 \right) .
 \end{align}
 The quantity $ \tilde{\Omega}_{m,n,l}$ is simply a convolution of BFKL gluon
Green's functions, originally defined in~\cite{Caporale:2015int}:
 \begin{align}\label{omega_mnl_tilde}
 &
 \tilde{\Omega}_{m,n,l}
 = 
 \int_0^{+\infty} dp_A \, p_A \int_0^{+\infty} dp_B \, p_B
 \int_0^{2\pi} d\phi_A \int_0^{2\pi} d\phi_B
 \\ & \nonumber
 \: \frac{e^{-im\phi_A} \, e^{il\phi_B} \,
       \left(p_A e^{i\phi_A}+k_1\right)^n \,
       \left(p_B e^{-i\phi_B}-k_2\right)^n}
      {
       \sqrt{\left(p_A^2+k_1^2+2 p_A k_1 \cos\phi_A\right)^n}
       \
       \sqrt{\left(p_B^2+k_2^2-2 p_B k_2 \cos\phi_B\right)^n}
      }
 \\ & \nonumber
 \varphi_m\left(|\vec{k_A}|,|\vec{p_A}|,Y_A-y_1\right)
  \varphi_l\left(|\vec{p_B}|,|\vec{k_B}|,y_2-Y_B\right) 
 \\ & \nonumber
 \varphi_n\left
   (\sqrt{p_A^2+k_1^2+2 p_A k_1 \cos\phi_A}
   ,\sqrt{p_B^2+k_2^2-2 p_B k_2 \cos\phi_B}
   ,y_1-y_2\right),
 \end{align}
where 
\begin{align}
 \varphi_{n} \left(|p|,|q|,Y\right) \; &= \; 
 \int_0^\infty d \nu   
 \cos{\left(\nu \ln{\frac{p^2}{q^2}}\right)}  
 \frac{e^{\bar{\alpha}_s  \chi_{|n|} \left(\nu\right) Y}}
      {\pi^2 \sqrt{p^2 q^2} }, \label{phin}
 \\
 \chi_{n} \left(\nu\right) \; &= \; 2\, \psi (1) - 
 \psi \left( \frac{1+n}{2} + i \nu\right) - 
 \psi \left(\frac{1+n}{2} - i \nu\right)
\end{align}
($\psi$ is the logarithmic derivative of Euler's gamma function).
 
As we mentioned previously, we would like to consider quantities that are easily measured
experimentally and moreover we want to eliminate as much as possible any dependence
on higher order corrections. Thus, we need to consider
ratios\footnote{See discussion in Refs.~\cite{Vera:2006un,Vera:2007kn}.} similar to Eq.~(\ref{Rmnlpqr})
which are defined on a partonic level though.
Therefore, in order to provide testable theoretical predictions 
against any current and forthcoming experimental data, 
we proceed in two steps. Firstly, we impose LHC kinematical cuts by  integrating ${\cal C}_{M N L}$ over the momenta of the tagged jets. More precisely,
\begin{align}
\label{Cmnl_int}
 & 
 C_{MNL} =
 \int_{Y_A^{\rm min}}^{Y_A^{\rm max}} \hspace{-0.25cm} dY_A
 \int_{Y_B^{\rm min}}^{Y_B^{\rm max}} \hspace{-0.25cm} dY_B
 \int_{k_A^{\rm min}}^{k_A^{\rm max}} \hspace{-0.25cm} dk_A
 \int_{k_B^{\rm min}}^{k_B^{\rm max}} \hspace{-0.25cm} dk_B
 \int_{k_1^{\rm min}}^{k_1^{\rm max}} \hspace{-0.25cm} dk_1
 \int_{k_2^{\rm min}}^{k_2^{\rm max}} \hspace{-0.25cm} dk_2
  \nonumber \\
 &
 \times  \delta\left(Y_A - Y_B - Y\right) {\cal C}_{MNL},
\end{align}
where the rapidity $Y_A$ of the most forward jet $k_A$ is restricted to $0 < Y_A < 4.7$
and the rapidity $Y_B$ of the most backward jet $k_B$ is restricted to $-4.7 < Y_B < 0$
while their difference $Y \equiv Y_A - Y_B$ is kept fixed at definite values 
within the range $6.5 < Y < 9$. Obviously, the last condition on the allowed values
of $Y$ makes both the integration ranges over $Y_A$ and $Y_B$ smaller than 4.7 units of rapidity.
Secondly, we remove the zeroth conformal spin contribution responsible for 
any collinear contamination (contributions that originate at $\varphi_0$)
and we minimise possible higher order effects by introducing the ratios
\begin{eqnarray}
\label{RPQMN}
R_{PQR}^{MNL} \, = \, \frac{C_{MNL}}{C_{PQR}}
\label{RmnlqprNew}
\end{eqnarray}
where $M, N, L, P, Q, R$ are positive definite integers.

Let us proceed now and present results for the ratios  $R_{PQR}^{MNL}(Y)$ in Eq.~(\ref{RmnlqprNew}) 
as functions of the 
rapidity difference $Y$ between the outermost jets for different momenta configurations and for two
center-of-mass energies: $\sqrt s = 7$ and $\sqrt s = 13$ TeV. 
For the transverse momenta $k_A$, $k_B$, $k_1$ and $k_2$ we impose the following cuts:
\begin{enumerate}
\item 
\begin{align}
\label{cut1}
&k_A^{min} = 35\, \text{GeV},\,\,\, k_A^{max} = 60\, \text{GeV}, \nonumber \\
&k_B^{min} = 45\, \text{GeV},\,\,\, k_B^{max} = 60\, \text{GeV}, \nonumber\\
&k_1^{min} = 20\, \text{GeV}, \,\,\,k_1^{max} = 35\, \text{GeV}, \nonumber\\
&k_2^{min} = 60\, \text{GeV}, \,\,\,k_2^{max} = 90\, \text{GeV}.
\end{align}
\item 
\begin{align}
\label{cut2}
&k_A^{min} = 35\, \text{GeV}, \,\,\,k_A^{max} = 60\, \text{GeV}, \nonumber \\
&k_B^{min} = 45\, \text{GeV}, \,\,\,k_B^{max} = 60\, \text{GeV}, \nonumber\\
&k_1^{min} = 25\, \text{GeV}, \,\,\,k_1^{max} = 50\, \text{GeV}, \nonumber\\
&k_2^{min} = 60\, \text{GeV}, \,\,\,k_2^{max} = 90\, \text{GeV}.
\end{align}
\end{enumerate}

To keep things simple, in both cuts, we set  $k_2$  to be larger
than all the other three jet momenta and we only vary the range of $k_1$.
In the cut defined in Eq.~(\ref{cut1}), $k_1$ is smaller that all the other
three jet momenta whereas in the cut defined in Eq.~(\ref{cut2}), the allowed
$k_1$ values overlap
with the ranges of $k_A$ and $k_B$. In the plots to follow, 
we plot the ratios for the cut defined in Eq.~(\ref{cut1}) with a red dot-dashed
line and the ratios for the cut defined in Eq.~(\ref{cut2}) with a blue dashed
line.

The numerical computation of the observables to be shown was done
in \textsc{Fortran}. \textsc{Mathematica} was used for various cross-checks.
We used the NLO MSTW 2008 PDF sets~\cite{MSTW:2009}  whereas,
regarding the strong coupling $\alpha_s$, a two-loop running coupling setup 
with $\alpha_s\left(M_Z\right)=0.11707$ was used.
\cod{Vegas}~\cite{VegasLepage:1978} 
as implemented in the \cod{Cuba} library~\cite{Cuba:2005,ConcCuba:2015}
was our main integration routine. We also made use of
 the library \cod{Quadpack}~\cite{Quadpack:book:1983} as well as
of a modified version 
of the \cod{Psi}~\cite{RpsiCody:1973} routine.
 
\begin{figure}[p]
\begin{center}
\vspace{-2cm}

   \includegraphics[scale=.38]{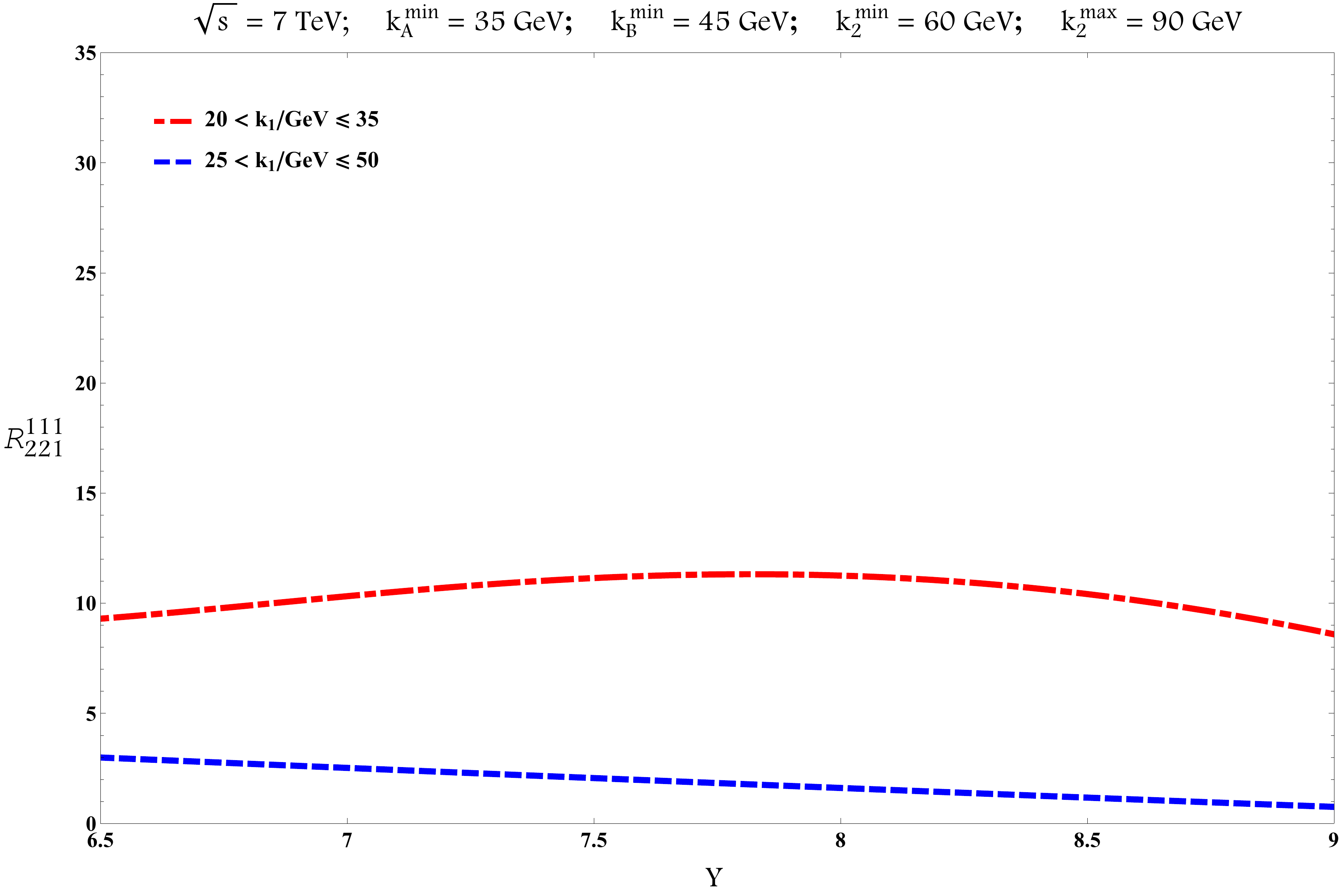}

   \vspace{.5cm}

   \includegraphics[scale=.38]{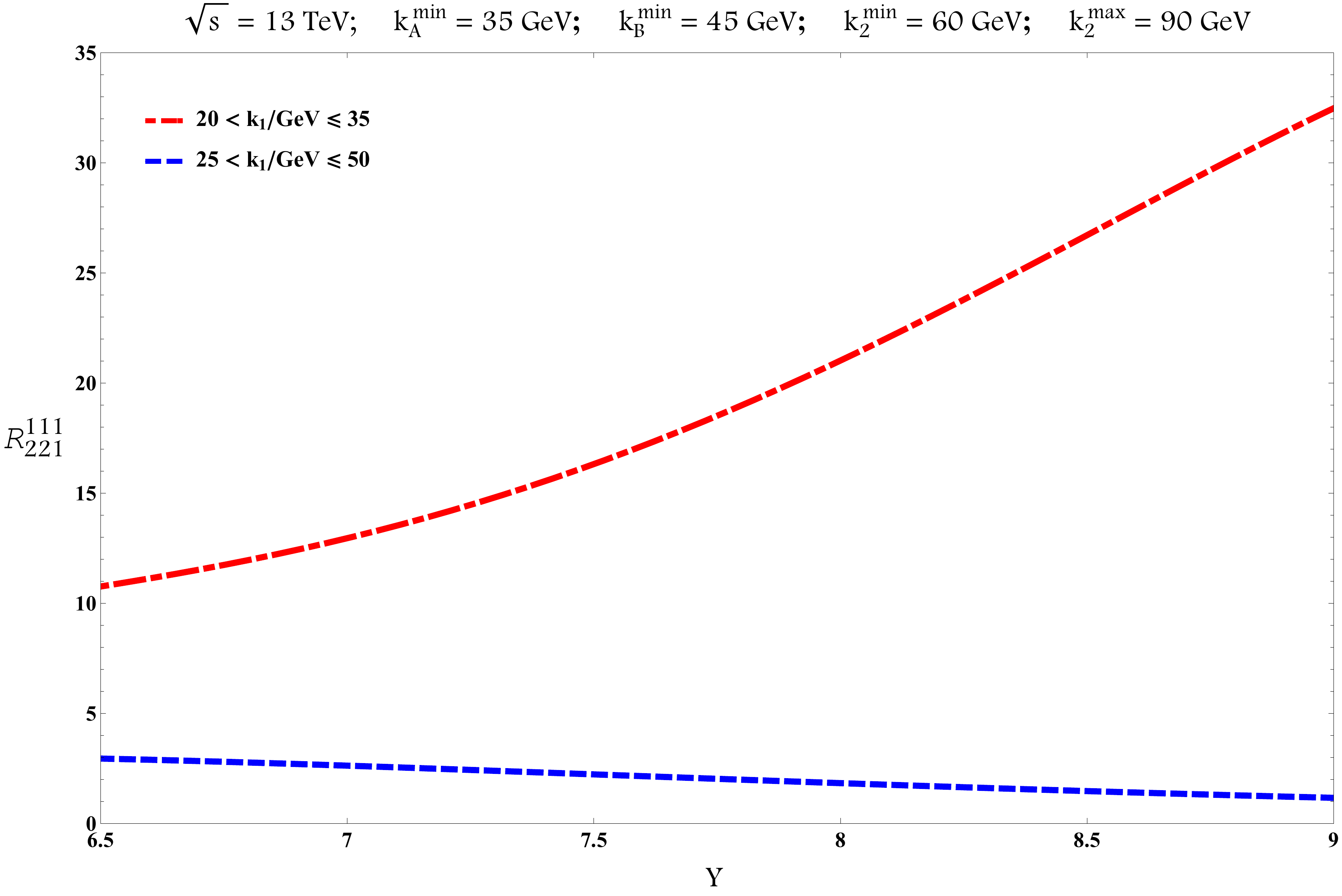}

\caption{\small $Y$-dependence of 
$R^{111}_{221}$
for $\sqrt s = 7$ TeV (top) and for $\sqrt s = 13$ TeV (bottom).} 
\label{fig:3}
\end{center}
\end{figure}

\begin{figure}[p]
\begin{center}
\vspace{-2cm}

   \includegraphics[scale=0.38]{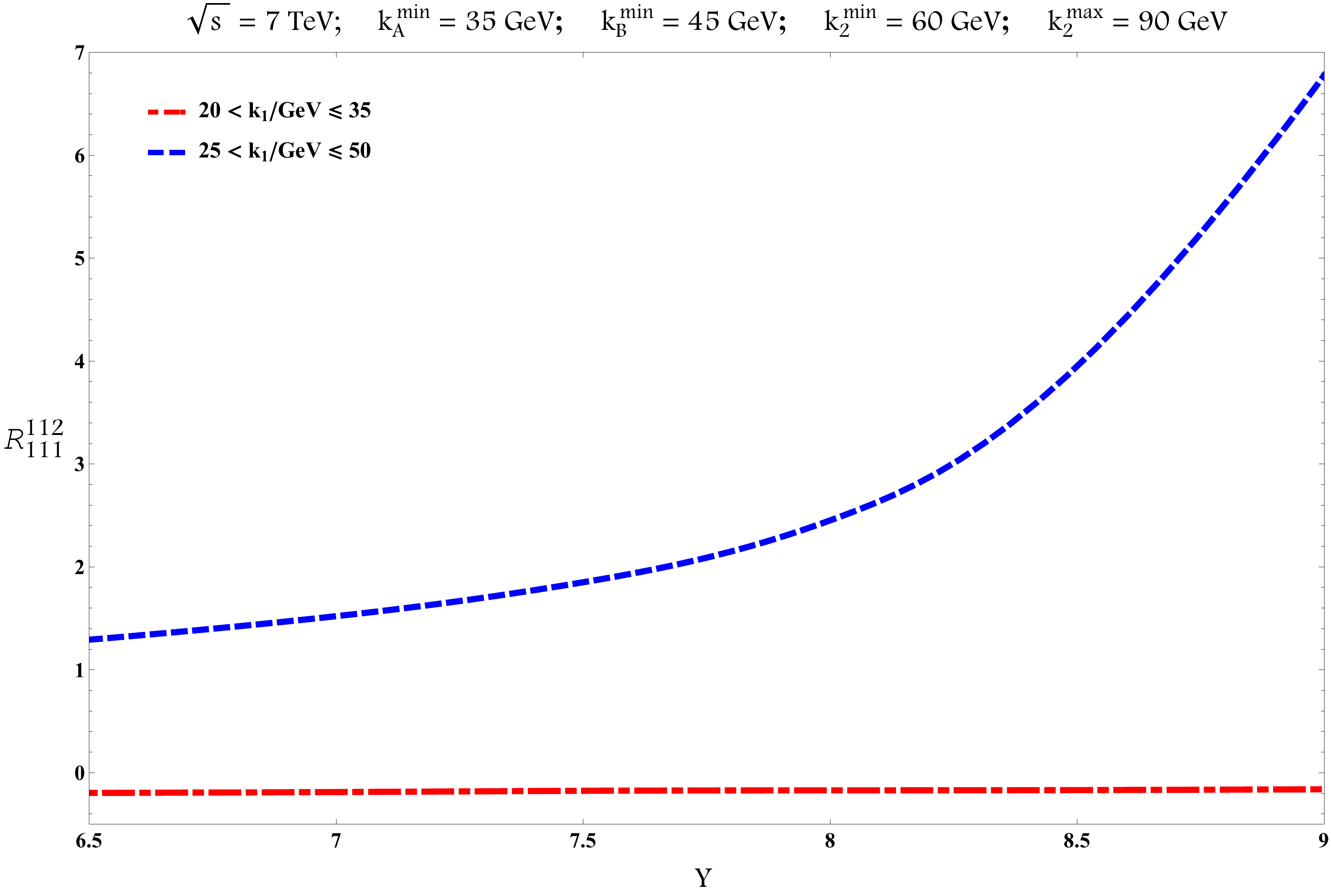}

   \vspace{.5cm}

   \includegraphics[scale=0.38]{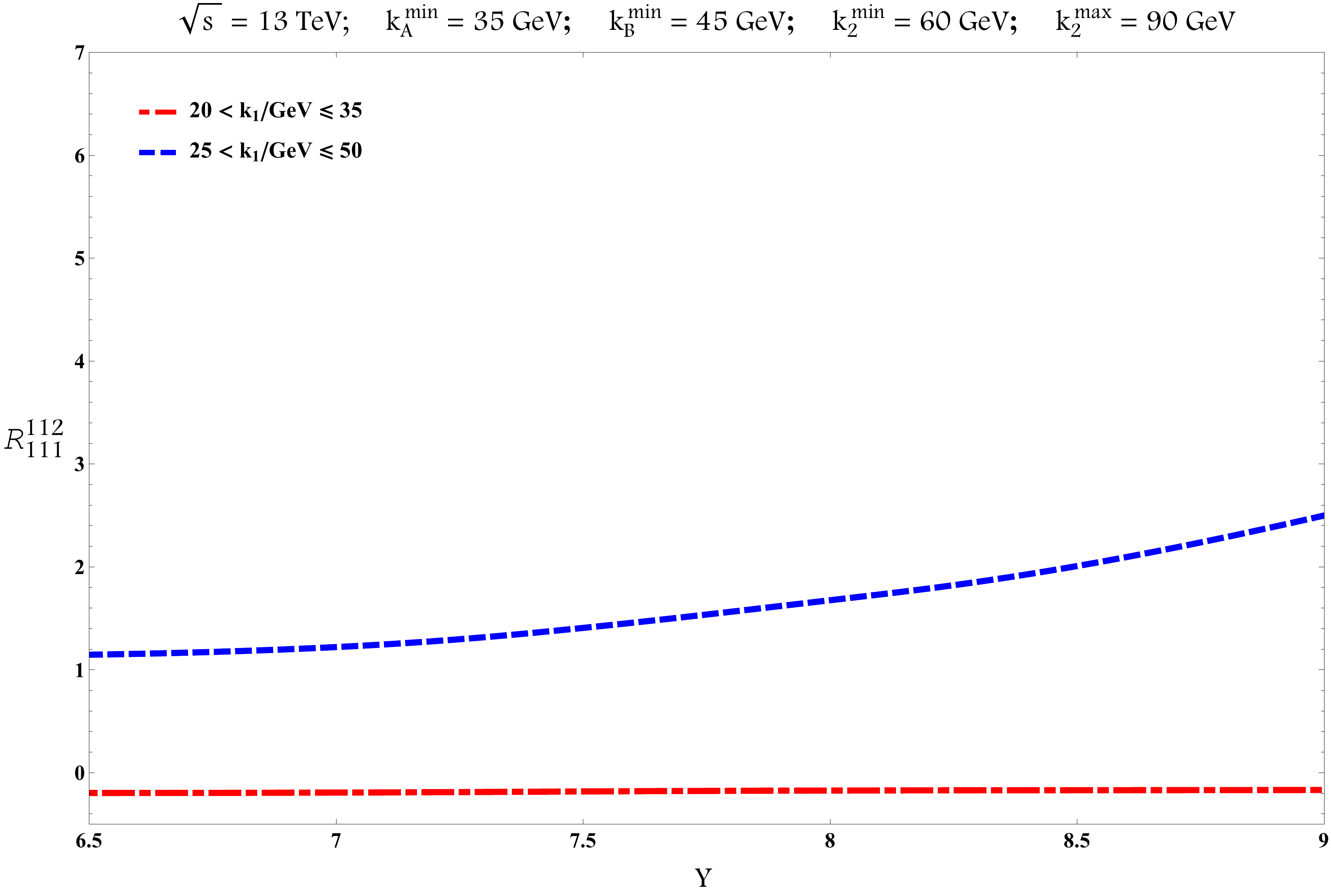}

\caption{\small $Y$-dependence of 
$R^{112}_{111}$
for $\sqrt s = 7$ TeV (top) and for $\sqrt s = 13$ TeV (bottom).} 
\label{fig:4}
\end{center}
\end{figure}

\begin{figure}[p]
\begin{center}
\vspace{-2cm}

   \includegraphics[scale=0.38]{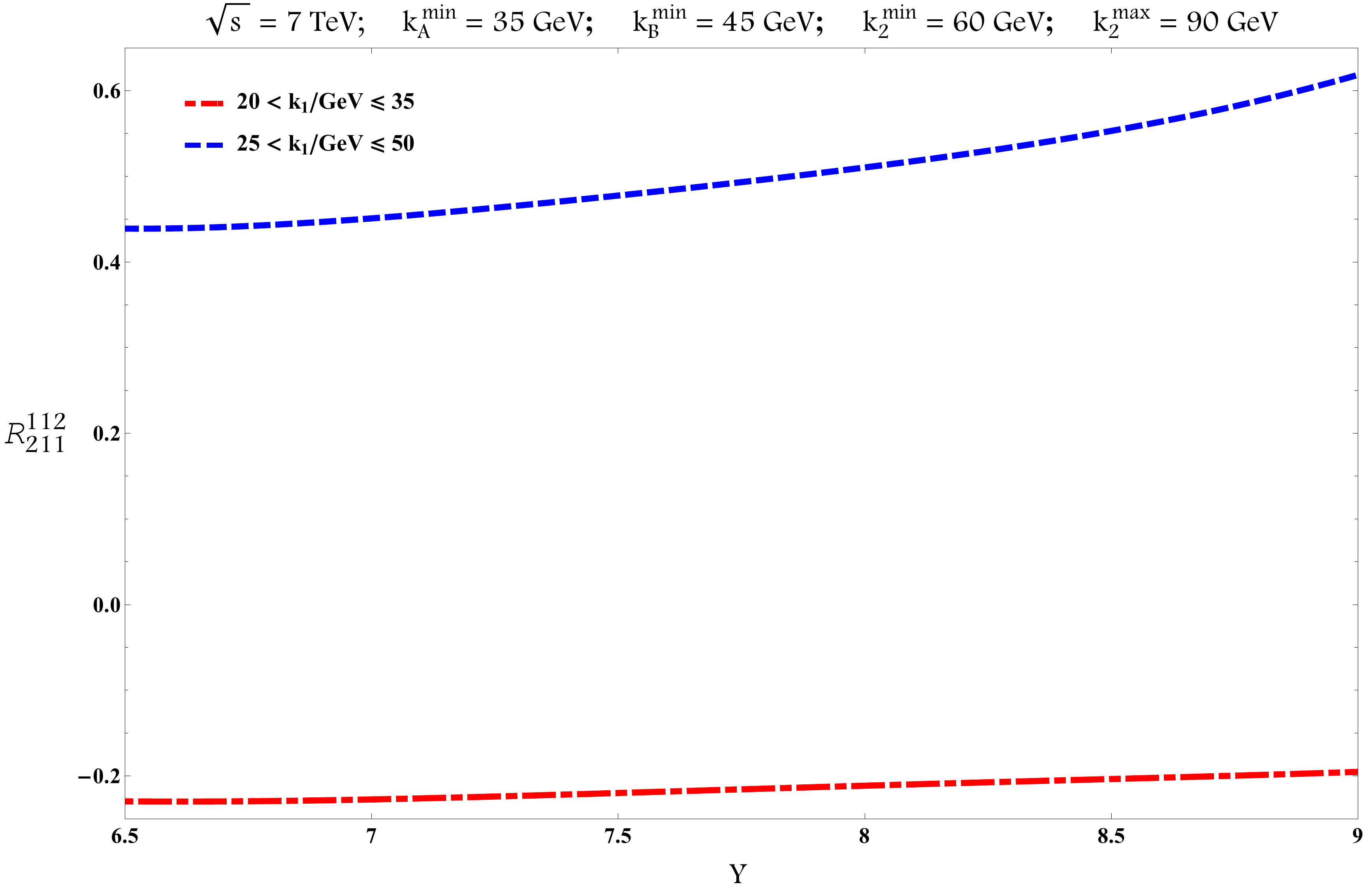}

   \vspace{.5cm}

   \includegraphics[scale=0.38]{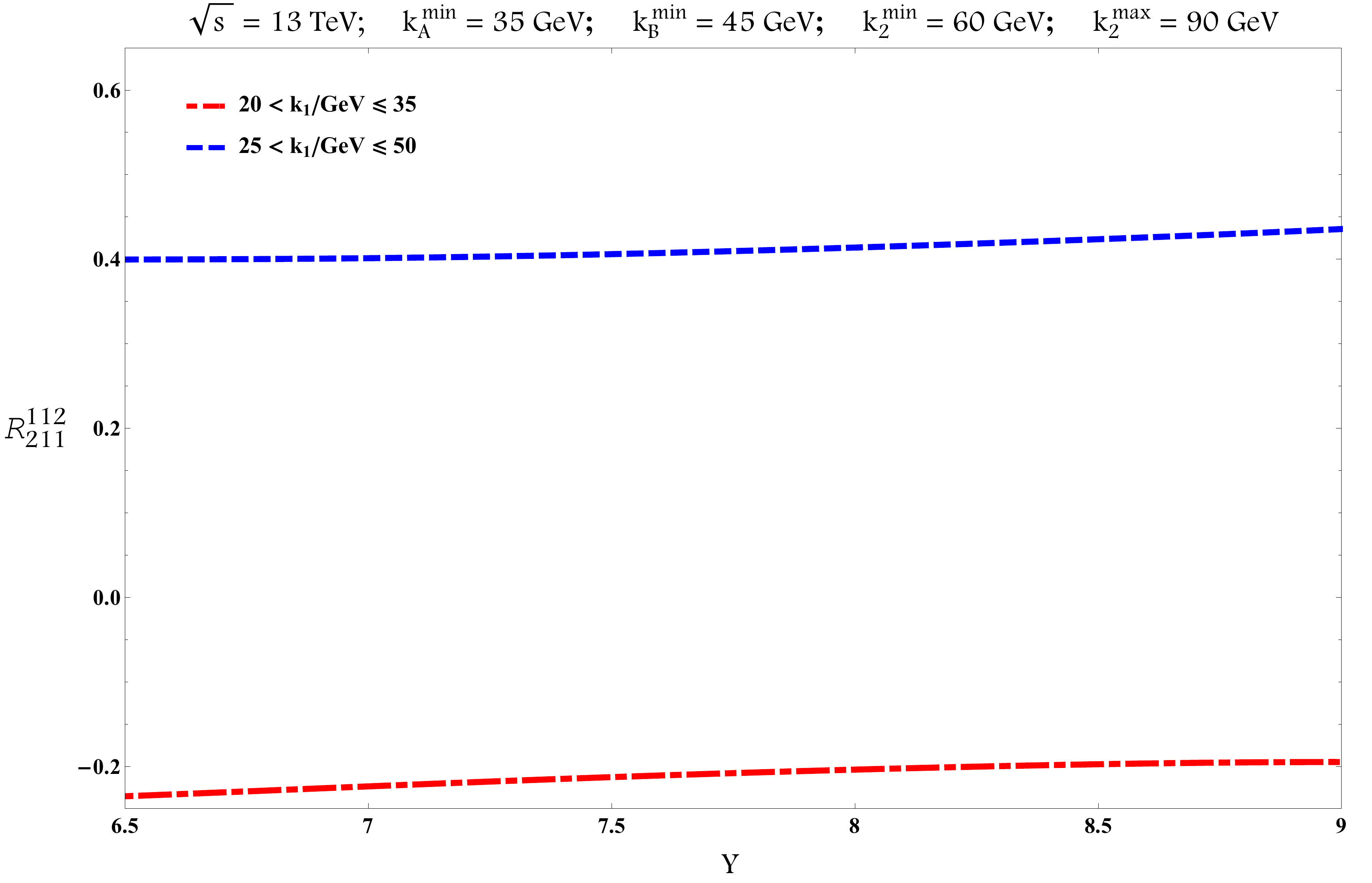}

\caption{\small $Y$-dependence of 
$R^{112}_{211}$
for $\sqrt s = 7$ TeV (top) and for $\sqrt s = 13$ TeV (bottom).} 
\label{fig:5}
\end{center}
\end{figure}

\begin{figure}[p]
\begin{center}
\vspace{-2cm}

   \includegraphics[scale=0.38]{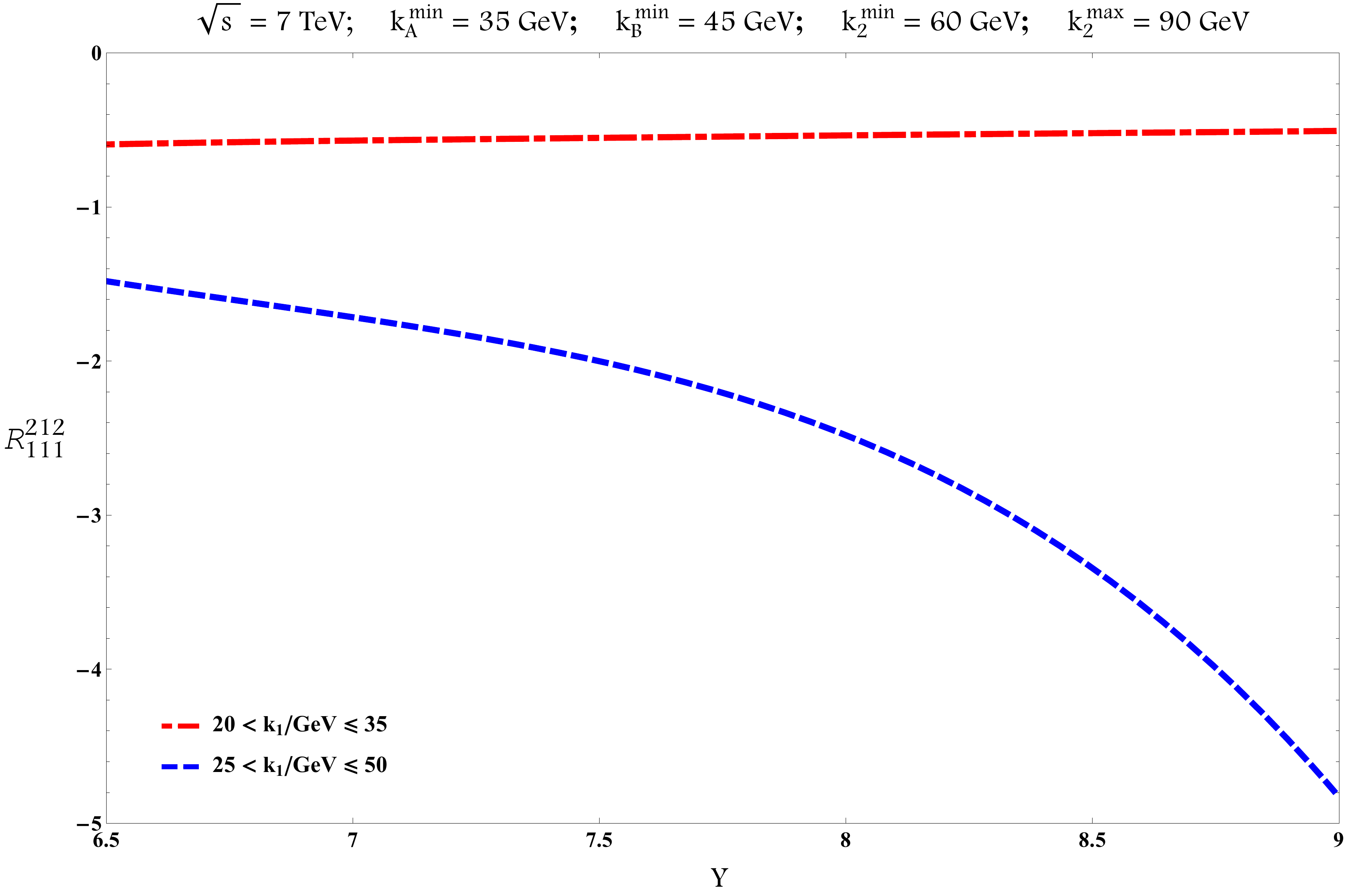}

   \vspace{.5cm}

   \includegraphics[scale=0.38]{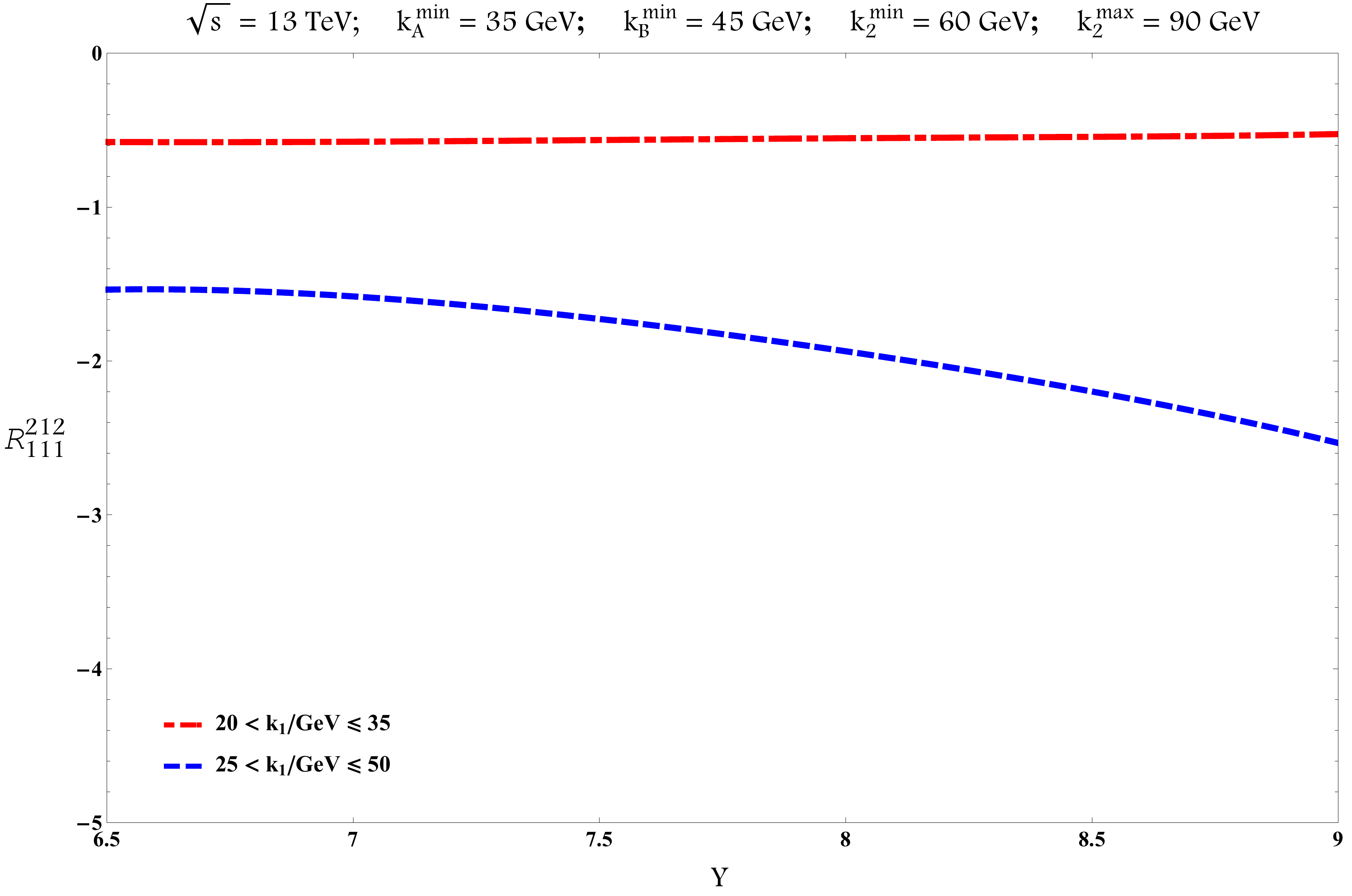}

\caption{\small $Y$-dependence of 
$R^{212}_{111}$
for $\sqrt s = 7$ TeV (top) and for $\sqrt s = 13$ TeV (bottom).} 
\label{fig:6}
\end{center}
\end{figure}

\begin{figure}[p]
\begin{center}
\vspace{-2cm}

   \includegraphics[scale=0.38]{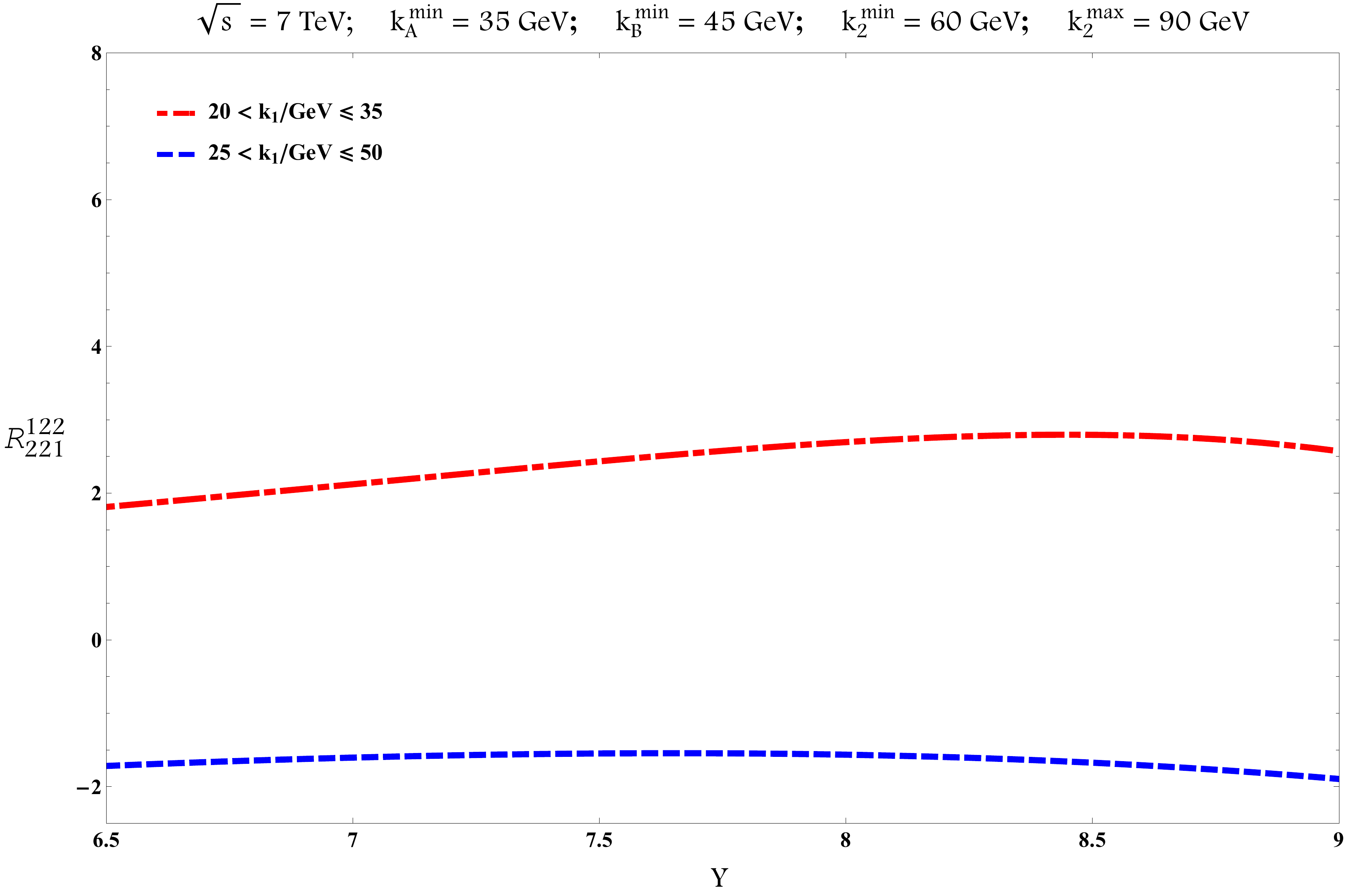}

   \vspace{.5cm}

   \includegraphics[scale=0.38]{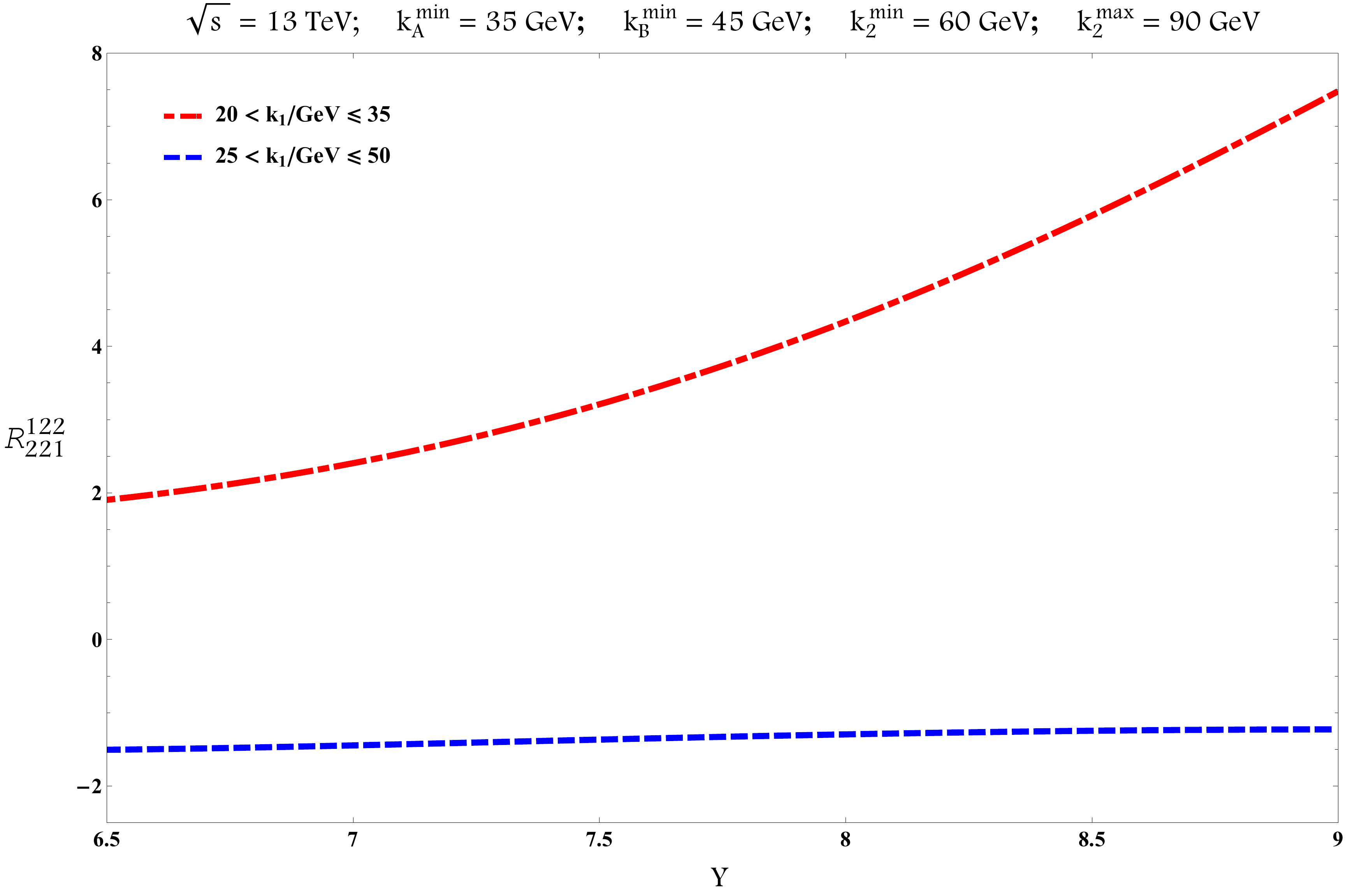}

\caption{\small $Y$-dependence of 
$R^{122}_{221}$
for $\sqrt s = 7$ TeV (top) and for $\sqrt s = 13$ TeV (bottom).} 
\label{fig:7}
\end{center}
\end{figure}

\begin{figure}[p]
\begin{center}
\vspace{-2cm}

   \includegraphics[scale=0.38]{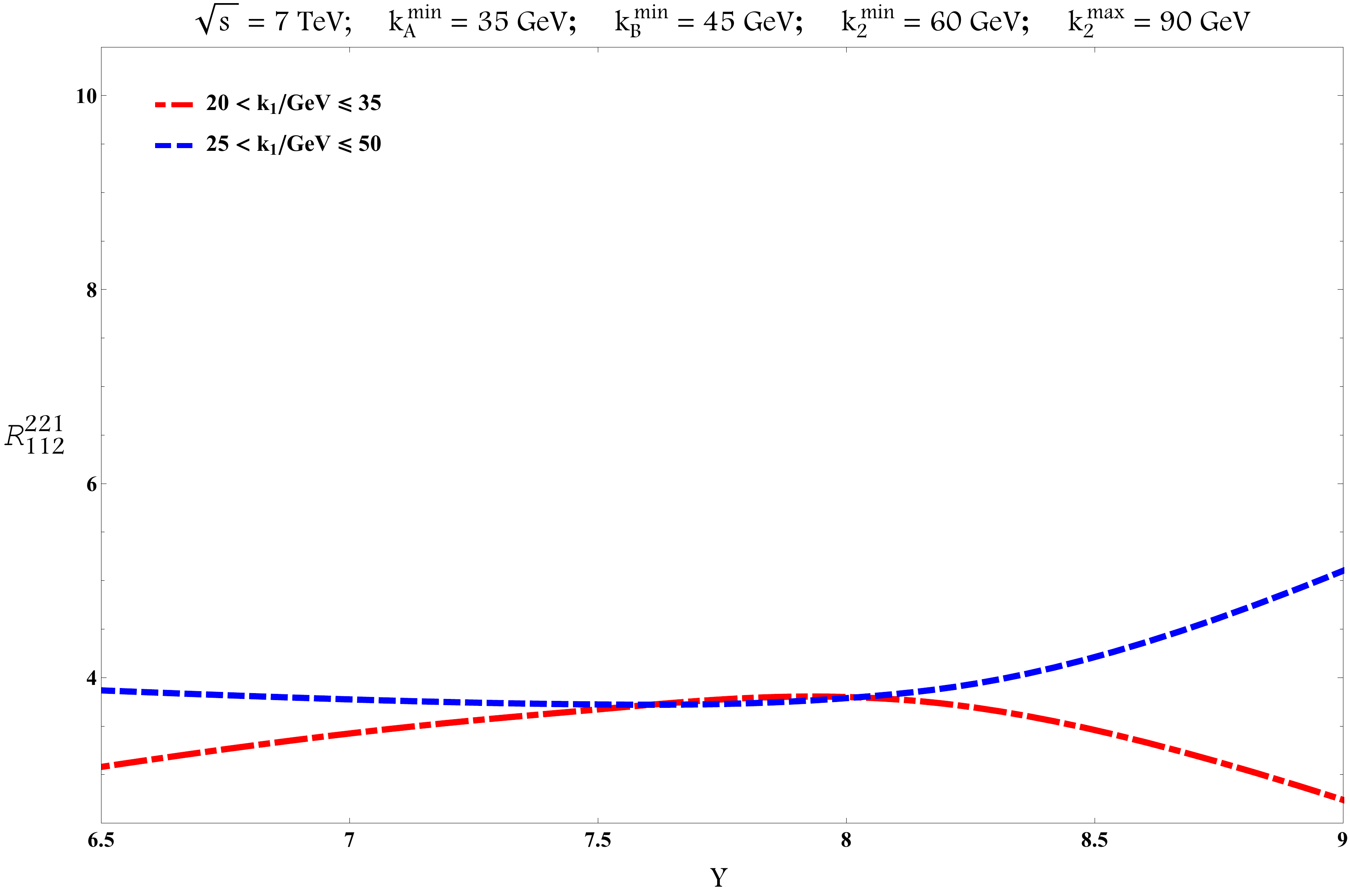}

   \vspace{.5cm}

   \includegraphics[scale=0.38]{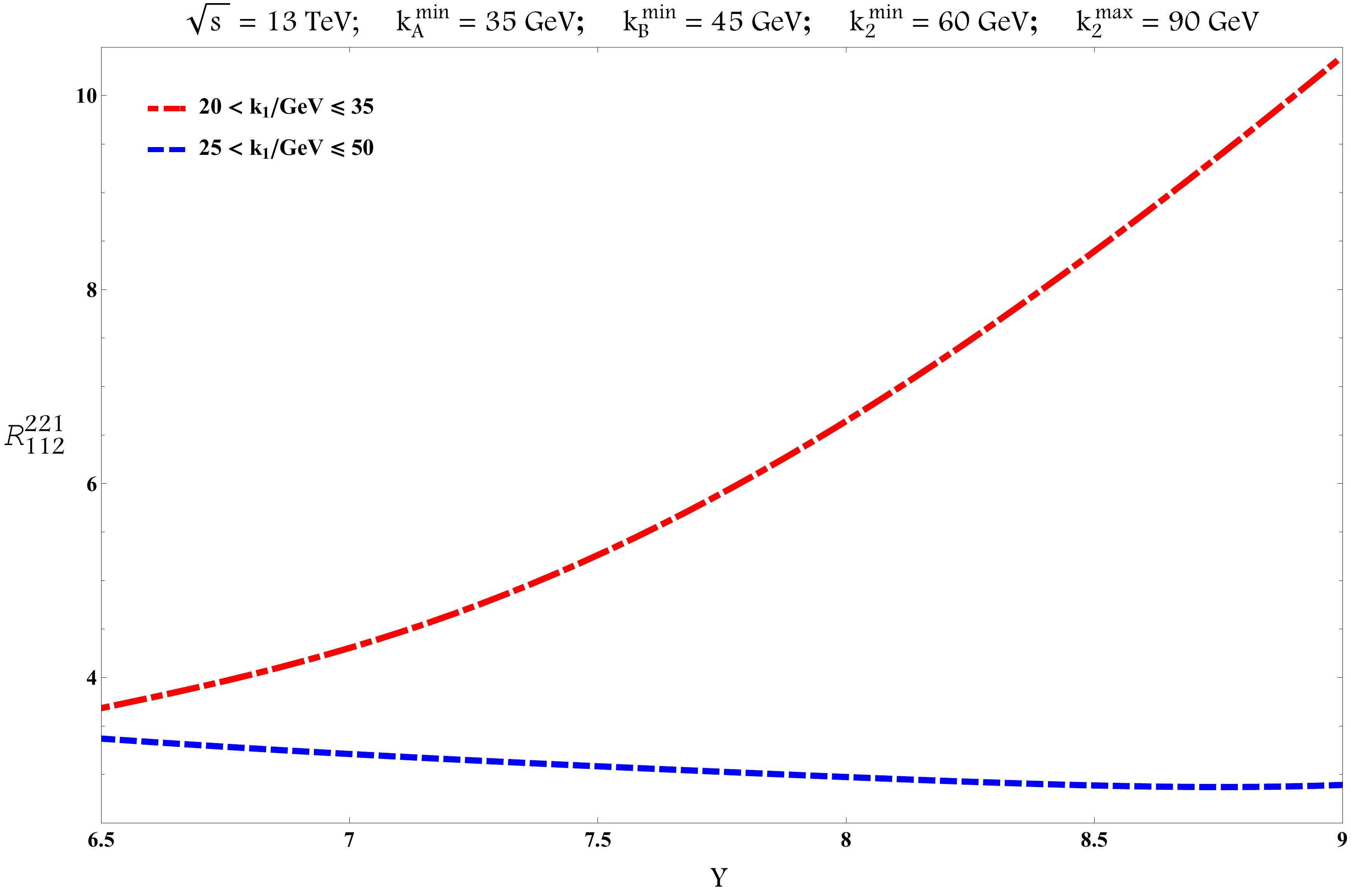}

\caption{\small $Y$-dependence of 
$R^{221}_{112}$
for $\sqrt s = 7$ TeV (top) and for $\sqrt s = 13$ TeV (bottom).} 
\label{fig:8}
\end{center}
\end{figure}

In the following, we present our results  for the
ratios $R^{111}_{221}$, $R^{112}_{111}$, $R^{112}_{211}$, $R^{212}_{111}$,
$R^{122}_{221}$, $R^{221}_{112}$
in Figs.~\ref{fig:3}$-$\ref{fig:8}. We place the $\sqrt s = 7$ TeV results
on the top of each figure and the $\sqrt s = 13$ TeV results 
at the bottom.

The functional dependence of the ratios $R^{MNL}_{PQR}$ on the rapidity difference between $k_A$ and $k_B$ is rather smooth. We can further notice that there are ratios  with an almost linear behaviour with $Y$ and with a rather small slope.  To be specific, the ratios represented by the blue curve in Fig.~\ref{fig:3} and the red curve in Figs.~\ref{fig:4},~\ref{fig:5} and~\ref{fig:6} demonstrate this linear behaviour in a striking fashion. Furthermore, whenever a ratio exhibits a linear dependence on $Y$ (for a certain kinematical cut of $k_1$) at colliding energy 7 TeV,  we observe that the ratio maintains almost the exact same linear behaviour --with very similar actual values-- at 13 TeV as well.

On the other hand, there are configurations for which the functional dependence on Y is much stronger and far from linear. In Fig.~\ref{fig:4}, the blue curve on the top rises from $\sim1.2$ at $Y=6.5$ to $\sim 6.8$ at $Y=9$, whereas in Fig.~\ref{fig:6} on the top it drops from $\sim (-1.5)$ to $\sim (-4.8)$ for the same variation in $Y$. Generally, if for some ratio there is a strong functional dependence on $Y$ for a $k_1$ of intermediate size (blue curve), this dependence is `softened' at higher colliding energy (see plots in Figs.~\ref{fig:4},~\ref{fig:5},~\ref{fig:6} and~\ref{fig:8}). However, for a $k_1$ of smaller size (red curve), we see that the functional dependence on $Y$ gets stronger at 13 TeV (Figs.~\ref{fig:3},~\ref{fig:7} and~\ref{fig:8}), unless of course it exhibits a linear behaviour as was discussed in the previous paragraph.

In all plots presented in Figs.~\ref{fig:3}$-$\ref{fig:8}, there is no red or blue curve that changes sign in the interval $6.5 < Y < 9$. Moreover, if a ratio $R^{MNL}_{PQR}$ is positive (negative) at 7 TeV, it will continue being positive (negative) at 13 TeV, disregarding the specific functional behaviour on $Y$.

In contrast to our main observation in Ref.~\cite{Caporale:2016soq} where in general, for most of the observables $R^{MN}_{PQ}$ there were no significant changes after increasing the colliding energy from 7 to 13 TeV, here we notice that, depending on the kinematical cut, an increase in the colliding energy may lead to a noticeable change to the shape of the functional Y dependence, e.g. red curve in Fig.~\ref{fig:3}, blue and red curve in Fig.~\ref{fig:8}. This is a very interesting point for the following reason. If a BFKL-based analysis for an observable dictates that the latter does not change much when the energy increases, this fact actually indicates that a kind of asymptotia has been reached, e.g. the slope of the gluon Green function plotted as a function of the rapidity for very large rapidities. In asymptotia, the dynamics is driven by pure BFKL effects whereas pre-asymptotic effects are negligible. In the present study, we have a mixed picture. We have ratios that do not really change when the energy increases and other ratios for which a higher colliding energy changes their functional dependence on Y. A crucial point that allows us to speak about pre-asymptotic effects, which in itself infers that BFKL is still the relevant dynamics, was outlined previously in this section: despite the fact that for some cases we see a different functional dependence on Y after raising the colliding energy, it is important to note that we observe no change of sign for any ratio $R^{MNL}_{PQR}$. Therefore, the four-jet ratio observables we are studying here are more sensitive to pre-asymptotic effects than the related three-jet ratio observables studied in Ref.~\cite{Caporale:2016soq}. Nevertheless, by imposing different kinematical cuts one can change the degree of importance of these effects. 

To conclude with, carefully combined choice of cuts for the $R^{MNL}_{PQR}$ observables and a detailed confrontation between theoretical predictions and data may turn out to be an excellent way to probe deeper into the BFKL dynamics.
 
\section{Summary \& Outlook}

We presented a first phenomenological study for some aspects of
the jets' azimuthal profile in LHC inclusive
four-jet production within the BFKL resummation framework.
Following up the work  in Ref.~\cite{Caporale:2015int}, where
a new set of BFKL probes was proposed for the LHC based on 
a partonic level study, 
we have calculated  
some of these observables here, after convoluting 
the previous results with
parton distribution functions and imposing LHC kinematical cuts,
at two different center-of-mass energies, $\sqrt s = 7, 13$ TeV. 

We have chosen an asymmetric kinematical cut 
with respect to the transverse momenta
of the most forward ($k_A$) and most backward ($k_B$) jet which is arguably 
a more interesting kinematical configuration that a symmetric cut since
it allows for an easier distinction between BFKL 
and fixed order predictions~\cite{Celiberto:2015dgl,Ducloue:2013wmi}.
The asymmetry was realised by imposing different lower limits
to $k_A$  and $k_B$ ($k_A^{min} = 35$ GeV and $k_B^{min} = 45$ GeV).
Additionally, we demanded for $k_2$ to be larger than both $k_A$ and $k_B$
whereas the value of the transverse momentum $k_1$ was allowed to be 
either smaller
than both $k_A$ and $k_B$ or overlapping the $k_A$ and $k_B$ range
of values.

We have plotted six generalised-azimuthal-ratio observables, 
$R^{111}_{221}$, $R^{112}_{111}$, $R^{112}_{211}$, $R^{212}_{111}$,
$R^{122}_{221}$, $R^{221}_{112}$,
as a function of the rapidity
distance $Y$ between $k_A$ and $k_B$ for $6.5 < Y <  9$.
A smooth functional dependence of the ratios on $Y$ appears to be the rule. 
It is noteworthy that 
the plots for ratios we presented exhibit in some cases considerable change when 
the colliding energy increases from 7 to 13 TeV. This tells us that
pre-asymptotic effects do play a role for the azimuthal ratios in inclusive
four-jet production.
A comparison with predictions for these observables from 
fixed order analyses as well as from the BFKL inspired 
Monte Carlo \cod {BFKLex}~\cite{Chachamis:2011rw,Chachamis:2011nz,Chachamis:2012fk,
Chachamis:2012qw,Caporale:2013bva,Chachamis:2015zzp,Chachamis:2015ico} 
seems to be the logical next step. Predictions from multi-purpose Monte Carlos tools should also be
pursued. 

We will conclude our discussion by stressing that it would be very interesting to have an
experimental analysis for these observables using existing and 
future LHC data.
We have the strong belief  that such
an analysis will be a big step forward
to the direction of gauging the applicability, at present energies, of the  BFKL dynamics 
in phenomenological studies.

\vspace{2cm}
\begin{flushleft}
{\bf \large Acknowledgements}
\end{flushleft}
GC acknowledges support from the MICINN, Spain, 
under contract FPA2013-44773-P. DGG acknowledges 
financial support from `la Caixa'-Severo Ochoa doctoral fellowship.
ASV and DGG acknowledge support from the Spanish Government 
(MICINN (FPA2015-65480-P)) and, together with FC and FGC, 
to the Spanish MINECO Centro de Excelencia Severo Ochoa Programme (SEV-2012-0249). 
FGC thanks the Instituto de F{\'\i}sica Te{\'o}rica 
(IFT UAM-CSIC) in Madrid for warm hospitality.

\end{document}